\documentclass[fleqn,apjl,usenatbib,  iop]{emulateapj}

\usepackage{comment}
\usepackage{ulem}
\usepackage{tikz}
\usepackage{color}
\usepackage{graphicx}
\usepackage{amsmath}
\usepackage{amsfonts}
\usepackage{outlines}
\usepackage{enumitem}

\usepackage[breaklinks,colorlinks,
   urlcolor=blue,citecolor=blue,linkcolor=blue]{hyperref}
\usepackage[all]{hypcap}

\newcommand{\beq}{\begin{equation}}
\newcommand{\eeq}{\end{equation}}
\newcommand{\bea}{\begin{eqnarray}}
\newcommand{\eea}{\end{eqnarray}}

\newcommand{\trm}[1]{\textrm{#1}}

\newcommand{\non}{\nonumber \\}

\DeclareGraphicsExtensions{.png,.pdf}

\begin{document}

\shortauthors{WEINBERG, SUN, ARRAS, \& ESSICK}
\shorttitle{TIDAL DISSIPATION IN WASP-12}

\title{Tidal dissipation in WASP-12}

\author{Nevin~N.~Weinberg$^{1}$,  Meng~Sun$^{2}$, Phil~Arras$^{2}$, and Reed~Essick$^{1}$}
\affil{$^1$Department of Physics, and Kavli Institute for Astrophysics and Space Research, Massachusetts Institute of Technology, \\Cambridge, MA 02139, USA\\
$^2$Department of Astronomy, University of Virginia, P.O. Box 
400325, Charlottesville, VA 22904-4325, USA}

\begin{abstract}
WASP-12 is a hot Jupiter system with an orbital period of $P= 1.1\trm{ day}$, making it one of the shortest-period giant planets known.  Recent transit timing observations by \citet{Maciejewski:16} and \citet{Patra:17} find a decreasing period with $P/|\dot{P}| = 3.2\trm{ Myr}$.  This has been interpreted as evidence of either orbital decay due to tidal dissipation or a long term oscillation of the apparent period due to apsidal precession.  Here we consider the possibility that it is orbital decay.  We show that the parameters of the host star are consistent with either a $M_\ast \simeq 1.3 M_\odot$ main sequence star or a $M_\ast \simeq 1.2 M_\odot$ subgiant.   We find that if the star is on the main sequence, the tidal dissipation is too inefficient to explain the observed $\dot{P}$. However, if it is a subgiant, the tidal dissipation is significantly enhanced due to nonlinear wave breaking of the dynamical tide near the star's center.  The subgiant models have a tidal quality factor $Q_\ast'\simeq 2\times10^5$ and an orbital decay rate that agrees  well with the observed $\dot{P}$.  It would also explain why the planet survived for $\simeq 3\trm{ Gyr}$ while the star was on the main sequence and yet is now inspiraling on a 3 Myr timescale.  Although this suggests that we are witnessing the last $\sim 0.1\%$ of the planet's life, the probability of such a detection is a few percent given the observed sample of $\simeq 30$ hot Jupiters in $P<3\trm{ day}$ orbits around  $M_\ast>1.2 M_\odot$ hosts.
\end{abstract}

\section{Introduction}\label{s:intro}

The orbits of hot Jupiters are expected to decay due to tidal dissipation within their host stars \citep{Rasio:96}. While there is considerable indirect evidence of orbital decay in the ensemble properties of hot Jupiter systems \citep{Jackson:08, Jackson:09, Hansen:10, Penev:12, Schlaufman:13, Teitler:14}, the recent transit timing observations of WASP-12 by \citet{Maciejewski:16} and \citet{Patra:17} could be the first direct evidence of orbital decay of an individual system.  They detect a decrease in the orbital period at a rate $\dot{P} =-29\pm 3\trm{ ms yr}^{-1}$. This corresponds to an inspiral timescale of just $P/|\dot{P}|=3.2\trm{ Myr}$ and a stellar tidal quality factor $Q_\ast'\approx 2\times 10^5$. 

As both studies note, it is difficult to tell whether the observed $\dot{P}$ is due to orbital decay or is instead a portion of a long-term ($\approx 14\trm{ yr}$) oscillation of the apparent period.  The latter could be due to apsidal precession if the eccentricity is $e\approx 0.002$. However, it is not clear how to maintain such an $e$ in the face of rapid tidal circularization.  \citet{Patra:17} mention gravitational perturbations from the star's convective eddies, a mechanism \citet{Phinney:92} proposed  to explain the small but nonzero eccentricities of pulsars orbiting white dwarfs. However, the host star ($M_\ast \simeq 1.3M_\odot$) has a very low mass convective envelope ($\simeq 10^{-3}M_\odot$) and we estimate that there is too little energy in the eddies to maintain an $e\sim 10^{-3}$.  Another mechanism that can cause decade-long oscillations of the period that \citeauthor{Patra:17} mention is the \citet{Applegate:92} effect, which invokes variations in the quadrupole moment of the star over a magnetic activity cycle.  However, \citet{Watson:10} estimate that for WASP-12b, this effect shifts the transit arrival times by $\Delta T \la 10\trm{ s}$ after $T\approx 10\trm{ yr}$.  This corresponds to an average $|\dot{P}| \simeq 2 P \Delta T /T^2  < 1 \trm{ ms yr}^{-1}$ \citep{Birkby:14}, more than an order of magnitude smaller than the measured value.

With a few more years of monitoring it should be possible to distinguish unequivocally between orbital decay and precession \citep{Patra:17}.  In this paper, we consider whether the decay explanation is plausible.  In Section \ref{sec:models}, we construct stellar models that fit the observed properties of WASP-12.  In Section \ref{sec:tides}, we describe the relevant tidal processes and then use the stellar models to calculate the expected rate of tidal dissipation. We conclude in Section \ref{sec:discussion}.

\section{Stellar Models of WASP-12}
\label{sec:models}

The WASP-12 host star has an effective temperature $T_{\rm eff}=6300\pm 150\trm{ K}$ and a mean density 
$\rho_\ast\equiv 3M_\ast / 4\pi R_\ast^3 =0.475 \pm 0.038 \trm{ g cm}^{-3}$ (\citealt{Hebb:09, Chan:11}; here and below we adopt the values from the latter reference).  Note that $\rho_\ast$ is measured solely from the transit parameters of the light curve (see \citealt{Seager:03}) and  is not derived from a fit to stellar evolution models, unlike the stellar mass $M_\ast$ and radius $R_\ast$.  The spectrum of WASP-12 is consistent with  a supersolar metallicity ($[\textrm{Fe/H}]=0.30\pm{0.10}$) and a spin that is slow ($v\sin i < 2.2\pm 1.5\trm{ km s}^{-1}$) and likely misaligned with the planet's orbital plane  \citep{Schlaufman:10, Albrecht:12}. By fitting stellar models to $T_{\rm eff}$, $\rho_\ast$, and the metallicity, \citeauthor{Chan:11} (2011; see also \citealt{Hebb:09, Enoch:10, Fossati:10, Maciejewski:11}) find $M_\ast = 1.36 \pm 0.14 M_\odot$, $R_\ast = 1.595\pm 0.071 R_\odot$, and a surface gravity $\log g_\ast = 4.164\pm 0.029$ (in cgs units).   Based on three separate age dating techniques (lithium abundance, isochrone analysis, and gyrochronology) \citet{Hebb:09} find that  WASP-12 is likely to be several Gyr old,  implying an age comparable to its main sequence lifetime.

We construct stellar models using the MESA stellar evolution code \citep{Paxton:11, Paxton:13, Paxton:15}, version 9575. We assume a solar abundance scale based on \citeauthor{Asplund:09} (2009; solar metallicity $Z=Z_\odot=0.0142$) and follow the MESA prescriptions given in \citet{Choi:16} for calculating the abundances, equation of state, opacity, and reaction rates.  

As we show below, the properties of WASP-12 are consistent with both $M_\ast \simeq 1.3 M_\odot$ main-sequence models and  $M_\ast \simeq 1.2 M_\odot$ subgiant models.  The range of subgiant models that fit the observations is sensitive to how convection and  mixing in radiative zones is implemented in MESA.  In particular, we find it is sensitive to the values of the parameters of mixing length theory $\alpha_{\rm MLT}$, overshooting $f_{\rm ov}$, semiconvection $\alpha_{\rm sc}$, and diffusive mixing.  Although recent studies are starting to place interesting constraints on some of these parameters \citep{Silva:11, Magic:15, Moravveji:15, Moravveji:16, Moore:16, Deheuvels:16}, there is still considerable uncertainty, especially as to how they depend on stellar mass, metallicity, and age. For simplicity, we therefore use the Schwarzschild criterion with $f_{\rm ov}=0$, we neglect diffusive mixing, and we consider a range of values for  $\alpha_{\rm MLT}$.

\begin{figure}
\centering
\vspace{-0.7cm}
\includegraphics[width=3.35in]{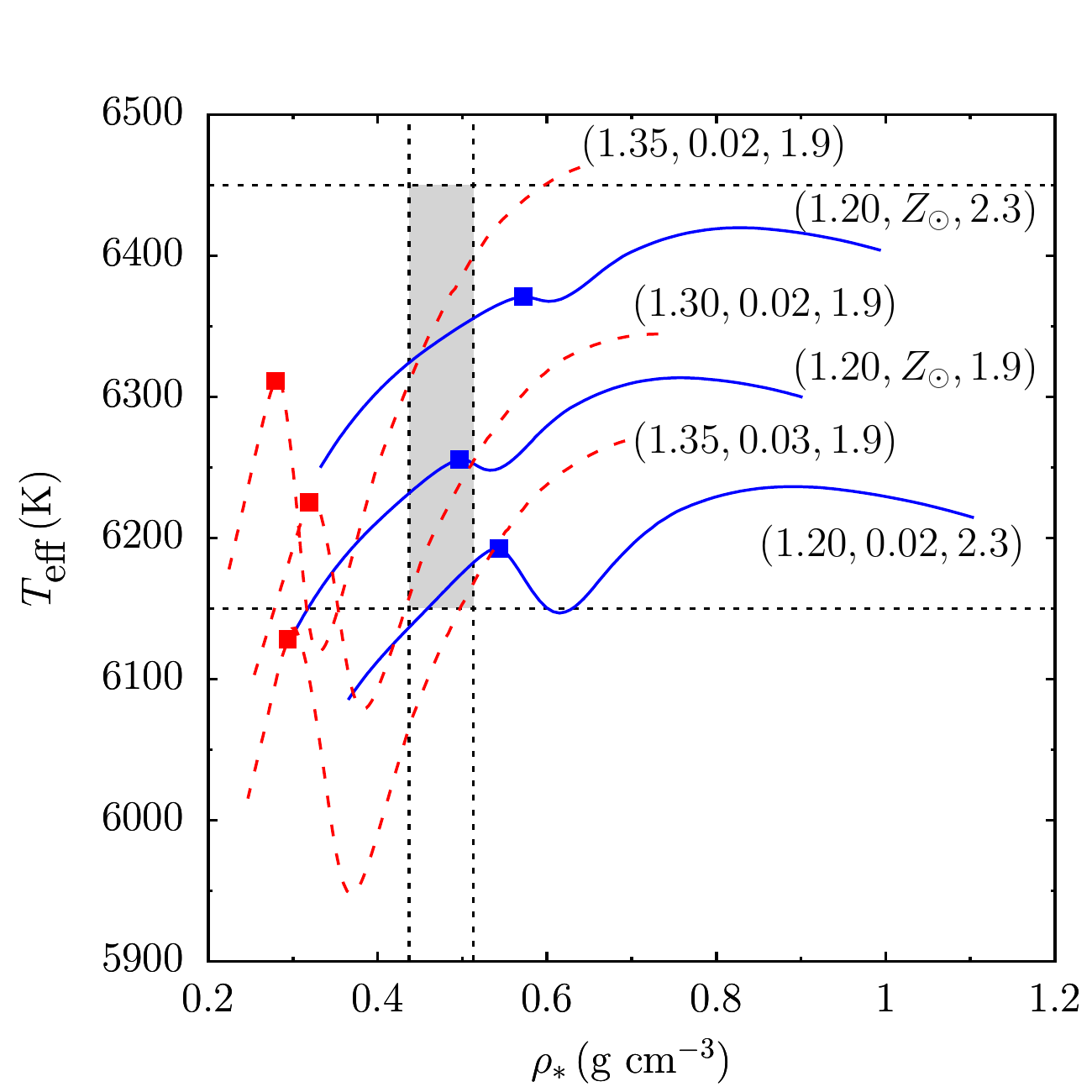} 
\caption{Evolution of the effective temperature $T_{\rm eff}$ and mean density $\rho_\ast$ for six stellar models.  Each model is labelled by $(M_\ast/M_\odot, Z, \alpha_{\rm MLT}$).  The evolution goes from right to left starting from when the star is $1\trm{ Gyr}$ old. The squares mark when the core ceases to be convective.  Observations of WASP-12 constrain its $T_{\rm eff}$ and $\rho_\ast$ to lie within the region indicated by the grey box.  The blue solid (red dashed) curves are models that match the observations when on the subgiant branch (main sequence). 
\label{fig:Trho_evolution}}
\end{figure}

In Figure \ref{fig:Trho_evolution} we show the evolution of $T_{\rm eff}$ and $\rho_\ast$ for six stellar models.  The three $M_\ast= 1.30 - 1.35 M_\odot$ models (red dashed curves) match the observed constraints (grey box) when the star is on the main sequence.  The three $M_\ast= 1.20 M_\odot$ models (blue solid curves) match the observed constraints during the post-main sequence phase, when the star is a subgiant and the core is no longer convective.  The different models are selected in order to illustrate that the evolution of $T_{\rm eff}$ and $\rho_\ast$ is sensitive to not only $M_\ast$, but also $Z$ and $\alpha_{MLT}$.  

All six models shown in Fig. \ref{fig:Trho_evolution} spend about $0.5\trm{ Gyr}$ within the measured range of $T_{\rm eff}$ and $\rho_\ast$.  During this portion of their evolution, the radii and surface gravity of the higher-mass models span $R_\ast=1.50-1.62 R_\odot$ and $\log g_\ast = 4.14-4.20$ while the lower-mass models span $R_\ast=1.47-1.55 R_\odot$ and $\log g_\ast = 4.13-4.18$. These are consistent with the (model-dependent) constraints reported in the literature.  

As we describe in Section \ref{sec:tides}, the efficiency of tidal dissipation is significantly enhanced if WASP-12 has a radiative core.  The only models with radiative cores that we find are consistent with the measured $T_{\rm eff}$ and $\rho_\ast$ are the subgiant models.   \citet{Torres:12} estimate a somewhat lower $T_{\rm eff}=6118\pm 64\trm{ K}$, which could match  the $T_{\rm eff}$ of  main sequence models with fully radiative cores (i.e., $M_\ast \la 1.1 M_\odot$). However, we find that such models have too high a $\rho_\ast$.   

\begin{figure}
\centering
\vspace{-0.7cm}
\includegraphics[width=3.35in]{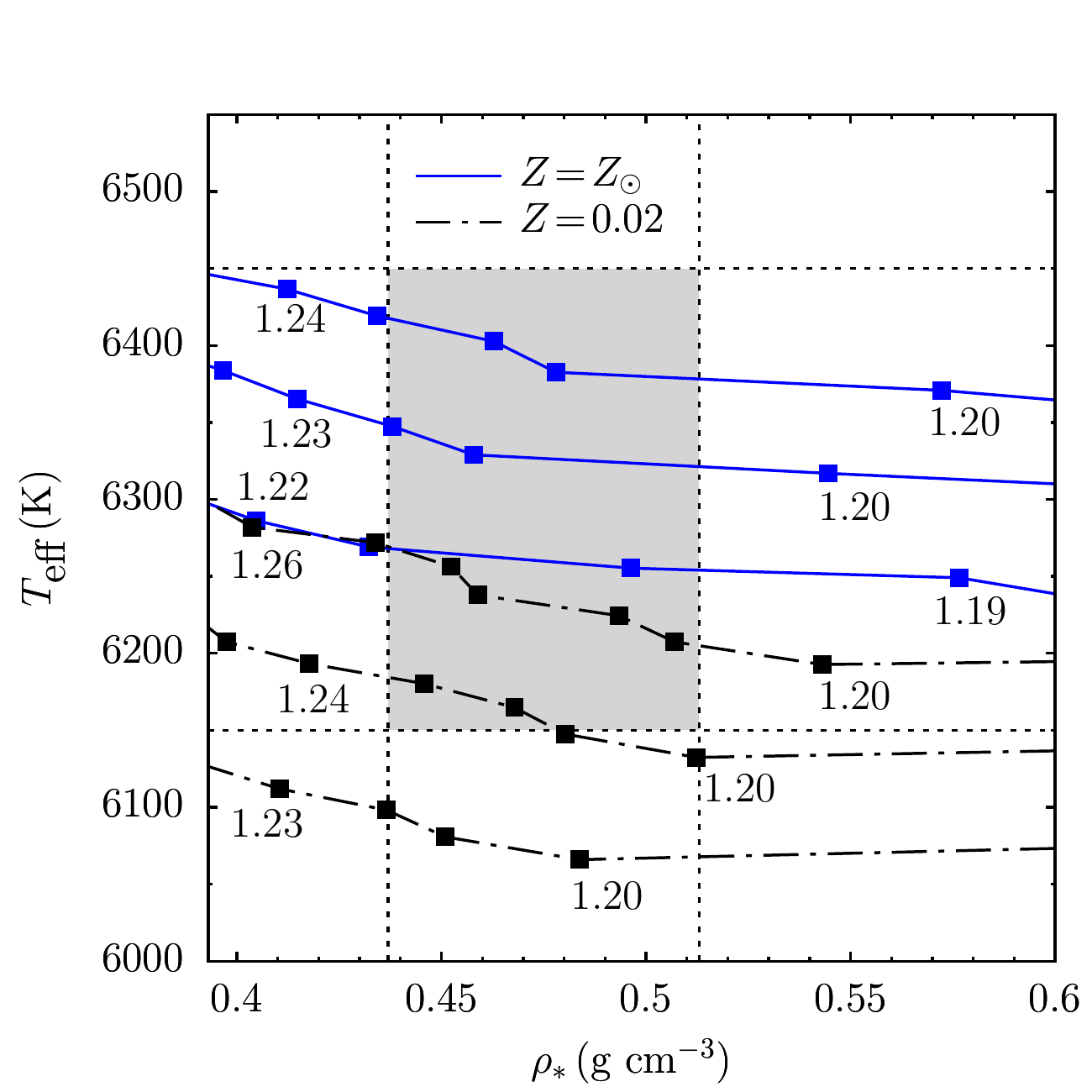} 
\caption{Effective temperature $T_{\rm eff}$ and mean density $\rho_\ast$ at the moment when the core ceases to be convective and the star enters the subgiant phase.  The labels indicate $M_\ast/M_\odot$ with points spaced by $0.01 M_\odot$ (connected by straight lines for clarity). The blue solid curves assume solar metallicity $Z=Z_\odot=0.0142$ and the black dashed-dotted curves assume $Z=0.02$.  The three curves for each $Z$ assume, from bottom to top, $\alpha_{\rm MLT}=1.9$, 2.1, and 2.3. Observations of WASP-12 constrain its $T_{\rm eff}$ and $\rho_\ast$ to lie within the region indicated by the grey box.
\label{fig:Trho_cc}}
\vspace{0.2cm}
\end{figure}

In Figure \ref{fig:Trho_cc} we show $T_{\rm eff}$ as a function of $\rho_\ast$ at the moment the core ceases to be convective and the star enters the subgiant phase.     We find that for a given $M_\ast$, increasing $\alpha_{\rm MLT}$ or decreasing $Z$ increases $T_{\rm eff}$ and $\rho_\ast$.  The models that are either inside or to the right (since $\rho_\ast$ decreases with age) of the grey box are consistent with the observations for a portion of the subgiant branch. The constraints are consistent with subgiant models whose parameters lie in the range $1.20 \la M_\ast/M_\odot \la 1.25$, $Z_\odot \la Z \la 0.03$ (i.e., $0\la[\textrm{Fe/H}]\la0.3$), and $1.9 \la \alpha_{\rm MLT} \la 2.3$.

\section{Tidal dissipation}
\label{sec:tides}

The orbit of WASP-12 appears circular ($e<0.05$; \citealt{Husnoo:12}) and, given the age of the system, the planet's rotation is expected to be synchronized \citep{Goldreich:66, Rasio:96}.  Therefore, any ongoing tidal dissipation must be occurring within the non-sychronized host star.  Dissipation mechanisms include turbulent damping of the equilibrium tide within the convective regions of the star and linear or nonlinear damping of the dynamical tide.  Studies of the former find $Q_\ast'\sim 10^8-10^9$ (\citealt{Penev:11}). This is more than three orders of magnitude too small a dissipation rate (too large a $Q_\ast'$) to explain the apparent orbital decay of WASP-12. We therefore focus on tidal dissipation due to the dynamical tide. 

The dynamical tide in WASP-12 is dominated by resonantly excited internal gravity waves.  Such waves propagate in the stratified regions of the star (where the Brunt-V\"ais\"al\"a buoyancy frequency $N^2>0$) and are evanescent within convective regions ($N^2<0$). As a result, the dynamical tide is excited near radiative-convective boundaries (RCBs), where its radial wavelength is long and it can couple well to the long lengthscale tidal potential \citep{Zahn:75, Zahn:77}.   

\begin{figure}
\centering
\vspace{-1.0cm}
\includegraphics[width=3.4in]{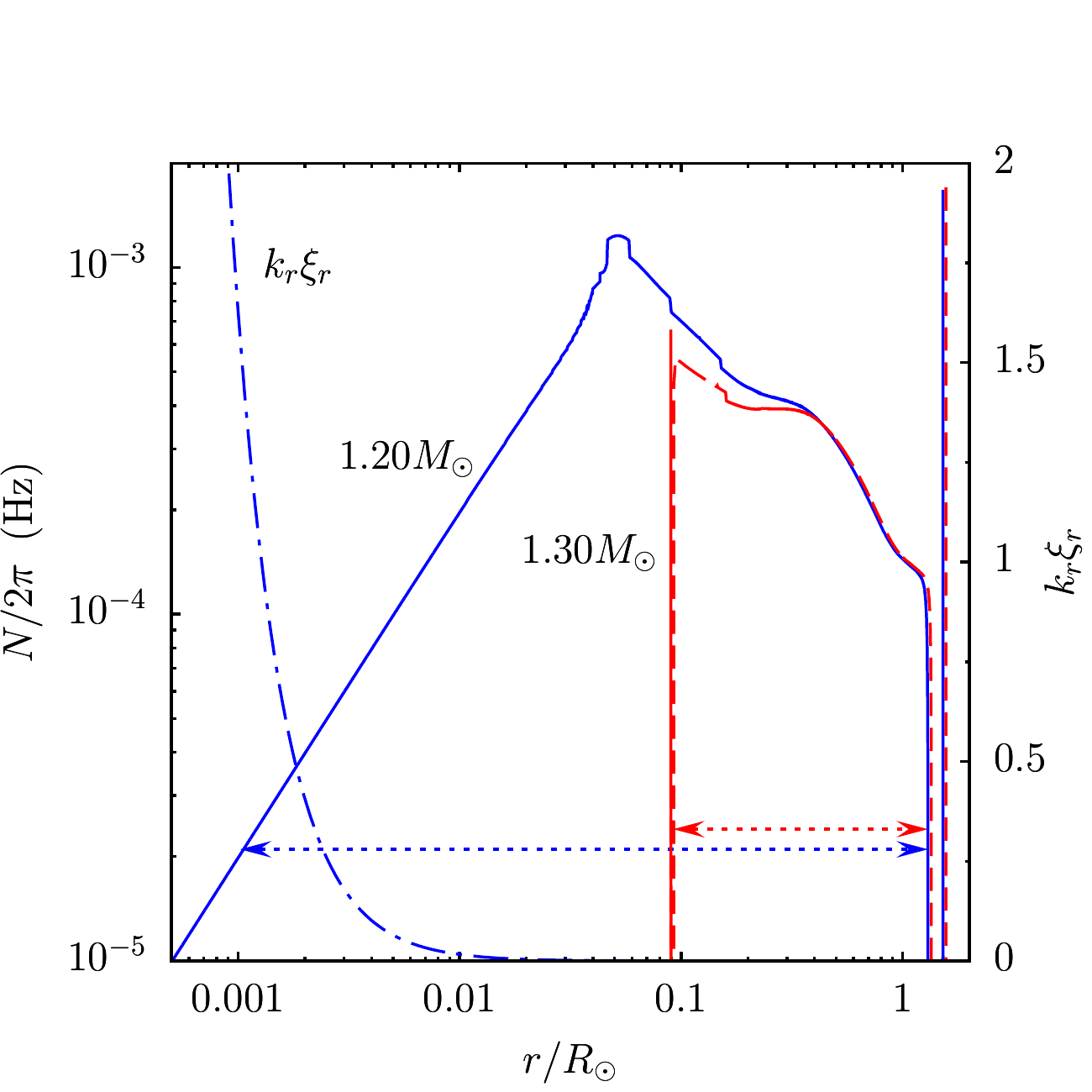} 
\caption{Radial profile of the Brunt-V\"ais\"al\"a frequency $N/2\pi$ (left-axis) and the nonlinearity measure $k_r\xi_r$ (right-axis).  The blue solid curve and red dashed curve show $N/2\pi$ for two WASP-12 models, respectively: the subgiant model $(M_\ast/M_\odot,Z,\alpha_{\rm MLT},\trm{Age/Gyr})$$=$$(1.20,Z_\odot,2.1, 3.7)$ and the main-sequence model $(1.30,0.02,1.9, 2.7)$.  The blue dashed-dotted curve shows $k_r\xi_r$ for the subgiant model.  The arrows indicate the turning points of the dynamical tide. 
\label{fig:brunt_profile}}
\vspace{0.07cm}
\end{figure}

When a star like WASP-12 (a late-F star) is on the main sequence, it has both a convective core and a convective envelope. When core hydrogen burning ends and the star evolves off the main-sequence and becomes a subgiant, its core ceases to be convective.  In Figure \ref{fig:brunt_profile} we show $N$ as a function of stellar radius $r$ for a main-sequence and subgiant model of WASP-12.  In the main-sequence model, the convective core extends from the center to $r\simeq  0.1 R_\odot$ and the convective envelope extends from $r \simeq 1.35 R_\odot$ to very near the surface.  The propagation cavity of the dynamical tide is determined by these two radii (they are its inner and outer turning points, respectively; see red arrows in Fig. \ref{fig:brunt_profile}).

In the subgiant model, by contrast, $N^2>0$ all the way to the center. We find a linear scaling with radius $N\simeq C r$ in the core, where $C \simeq 0.1(R_\odot \trm{ s})^{-1}$.  The dynamical tide propagates where the tidal frequency $\omega < N(r)$; for the dominant $\ell=2$ tide, $\omega=2\Omega$, where $\Omega$ is the orbital frequency. Thus, the tide raised by WASP-12b ($\omega/2\pi = 21.2\trm{ $\mu$Hz}$) has an inner turning point at $r\simeq \omega/C  \simeq 10^{-3} R_\odot$ during the subgiant phase (blue arrows in Fig. \ref{fig:brunt_profile}).  The dynamical tide propagates much closer to the center of the star when the star is a subgiant compared to when it is on the main sequence.

\subsection{Dynamical tide luminosity and wave breaking}

If the dynamical tide loses very little energy in the group travel time between turning points, it forms a global standing wave. Conversely, if it loses a significant fraction of its energy between turning points, it behaves more like a traveling wave excited near the outer convection zone and traveling inward to the center.  We will show that the dynamical tide is a standing wave for the main sequence models of WASP-12 and a traveling wave for the subgiant models.  We now calculate the dynamical tide luminosity $L$ assuming a traveling wave.

In the gravity wave propagation zone, the traveling wave luminosity is given by $L (r)= r^2 \int d\Omega\, \rho \psi_{\rm dyn} \dot{\xi}_{r,\rm dyn}$, where $\psi=\delta p/\rho + U$, $\delta p$ is the Eulerian pressure perturbation, $U$ is the tidal potential, $\xi_r$ is the radial displacement, and the subscript ``dyn" denotes the short-wavelength, dynamical tide piece. $L(r)$ is nearly constant with $r$ in the propagation zone (except near sharp features such as density variations on short lengthscales\footnote{Our stellar models have one or two sharp spikes in $N$ (near $\sim 0.1 R_\odot$) due to composition discontinuities that form as the convective core shrinks.  However, these spikes are unphysical; we find that they disappear when we include overshooting and diffusive mixing.  Here we simply smooth over them in order to calculate $L$.}).   To compute $\xi_r$ and $\psi$, we solve the equations of motion of the linear tide (e.g., \citealt{Weinberg:12}). We use the Cowling approximation, in which the perturbed gravity is ignored, a good approximation for the short-wavelength dynamical tide. A mechanical boundary condition $\psi - U = g \xi_r$ is used at the surface of the star, and the inward-going traveling wave boundary condition $d(\psi-\psi_0)/dr = i k_r (\psi - \psi_0)$ is applied at a radius well within the propagation zone. Here $\psi_0$ is an approximation of the long-wavelength, particular solution, called the ``finite frequency equilibrium tide" (see \citealt{Arras:10}); it is given by $\Lambda^2\psi_0=\omega^2d(r^2\xi_{r,\rm eq})/dr$, where  $\Lambda^2=\ell(\ell+1)$ and $\xi_{r,\rm eq}=-U/g$ is the radial displacement of the zero-frequency equilibrium tide.  The dynamical tide piece of the solution is given by $\xi_{r,\rm dyn}=\xi_r -\xi_{r,\rm eq}$ and $\psi_{\rm dyn}=\psi-\psi_0$.

This numerical calculation of $L$ may be compared to analytic treatments in which approximate solutions in the radiative and  convection zones are matched across the RCB \citep{Zahn:75, Goldreich:89, Goodman:98, Kushnir:17}. While an analytic treatment is, in principle, useful in providing simple formulae,  the solution in the convection zone and the matching conditions at the RCB are complicated and can depend on $\omega$ and the size of the outer convection zone. Nonetheless, we motivate an approximate fitting formula as follows. Given an equilibrium tide displacement $\xi_{r,\rm eq} \simeq -U/g$, the dynamical tide near the RCB is $\xi_{r,\rm dyn} \simeq \zeta (\lambda/r) \xi_{r,\rm eq}$, where $\zeta$ is a dimensionless constant that depends on the structure of the convection zone found by the matching conditions, and $\lambda = [- (\Lambda^2 /\omega^2 r^2)dN^2/dr]^{-1/3}$ is the wavelength near the RCB. By the continuity equation, $\Lambda^2 \psi_{\rm dyn}/\omega^2 r^2 \simeq d\xi_{r,\rm dyn}/dr \simeq \xi_{r,\rm dyn}/\lambda$. Thus,  for the dominant $\ell=2$ gravity wave,
\bea
L & = & A_L \frac{GM_p^2}{r_c} \left( \frac{r_c}{a} \right)^6 \left( \frac{\rho_c}{\bar{\rho}_c} \right) \left( \frac{\omega}{\omega_c} \right)^{8/3} \omega
\non &\simeq& 7\times10^{29}A_L \left(\frac{M_p}{10^{-3} M_\ast}\right)^2 \left(\frac{M_c}{M_\odot}\right)^{-7/3}  \left(\frac{r_c}{R_\odot}\right)^{12}
\non &&\times \left(\frac{\rho_c}{10^{-3}\trm{ g cm}^{-3}}\right) \left(\frac{P}{\trm{day}}\right)^{-23/3}\trm{ erg s}^{-1},
\label{eq:L}
\eea
Here $a$ is the semi-major axis, $r_c$ is the radius of the RCB where the wave is excited, $\rho_c$ is the density at $r_c$, $\bar{\rho}_c = 3M_c/4\pi r_c^3$ and $M_c$ are the mean density and enclosed mass within $r_c$, and $\omega_c = (GM_c/r_c^3)^{1/2}$ is the dynamical frequency at $r_c$. The  dimensionless prefactor $A_L \simeq 0.02\zeta^2[-(r_c/\omega_c^2)dN^2/dr]^{-1/3}$. 

Equation (\ref{eq:L}) is similar to the form derived by \citet{Kushnir:17}. It is useful if $A_L$ is nearly constant for different $P$ and stellar models.  In practice, we find that this is not the case.  Specifically, we find that for $P\la 2 \trm{ day}$, $A_L$ increases as $P$ decreases (this is because at such large forcing frequencies, the wavelength is not sufficiently small compared to a scale height near the RCB; see \citealt{Barker:11b}). Furthermore, at a fixed $P=1.1\trm{ day}$, 
we find that the different WASP-12 subgiant models give values in the range $0.2 \la A_L \la 0.6$.  Because of the complicated behavior of $A_L$, we rely on the numerical calculation of $L$ rather than Eq. (\ref{eq:L}).

If a fraction $\eta$ of the wave luminosity $L$ is deposited in a single group travel time across the star, then 
\citep{Goldreich:66, Ogilvie:14}
\bea
\frac{P}{|\dot{P}|}&=&
\frac{G^{2/3}M_pM_\ast \Omega^{2/3}}
{3(M_p+M_\ast)^{1/3}\eta L}
\non &\simeq&
\frac{9.1}{\eta L_{30}} \left(\frac{M_\ast}{M_\odot}\right)^{2/3}\left(\frac{M_p}{M_{\rm Jup}}\right)\left(\frac{P}{\trm{day}}\right)^{-2/3}\trm{ Myr},
\hspace{0.3cm}
\label{eq:PPdot}
\eea
where $L_{30}=L/10^{30} \trm{ erg s}^{-1}$. 
The value of $\eta$ depends on how efficiently the dynamical tide is dissipated as it propagates through the radiative interior. 

The principal dissipation mechanisms acting on the dynamical tide are damping due to radiative diffusion and nonlinear wave interactions \citep{Goodman:98, Barker:10, Weinberg:12, Essick:16, Chernov:17}. Radiative damping at a rate $\gamma$ causes the amplitude of the tide to decrease by a factor of $\exp(-\gamma t_{\rm gr})$ in a group travel time $t_{\rm gr}$ across the star. Damping due to nonlinear interactions is especially strong if the wave displacement $\xi_r$ is so large that $k_r \xi_r \ga 1$, where $k_r \simeq \Lambda N / \omega r$ is the radial wavenumber. Such a strongly nonlinear wave overturns the local stratification and breaks.  Since it deposits all of its energy and angular momentum before reflecting, wave breaking implies $\eta\simeq 1$ \citep{Barker:10, Barker:11b}.

We can estimate $k_r \xi_r$ in the WKB approximation using conservation of energy flux, which states that $\rho N^2 v_{\rm gr} |\xi_r|^2 \simeq L/4\pi r^2$ as the dynamical tide  propagates inward from the envelope RCB  (\citealt{Goodman:98}). Here $v_{\rm gr}\simeq \omega/k_r$ is the radial group velocity  and $\xi_r$ now denotes the rms radial displacement averaged over time and angle at fixed radius.  This gives
\bea
k_r \xi_r &\simeq& \sqrt{\frac{\Lambda^3 N L}{4\pi \rho r^5 \omega^4}}
\non &\simeq&1.3 \left(\frac{C_{0.1} L_{30}}{\rho_2}\right)^{1/2} \left(\frac{P}{\trm{day}}\right)^{2}  \left(\frac{r}{10^{-3}R_\odot}\right)^{-2}.
\hspace{0.2cm}
\label{eq:krxir}
\eea
The second line represents the scaling in a radiative core, where $N=Cr$ with $C_{0.1}=C/0.1 (\trm{$R_\odot$ s})^{-1}$ and  $\rho_2 = \rho /10^2\trm{ g cm}^{-3}$.  Our numerical solutions of $k_r\xi_r$ agree well with this WKB estimate.

\subsection{Resonance locking}
As a star evolves, its $g$-mode frequencies can increase, allowing them to sweep into resonance with the tidal frequency. If the resulting tidal torques are sufficiently large, the dynamical tide can end up in a stable ``resonance lock" and drive orbital decay on the stellar evolution timescale. Resonance locking has been invoked to explain the observed properties of a variety of tidally interacting binaries \citep{Witte:02, Fuller:12, Burkart:13, Burkart:14, Fuller:16, Fuller:17a}. 

We find that resonance locking cannot explain the apparent orbital decay of WASP-12. This is because the $g$-mode frequencies in the models evolve too slowly for a mode to remain in resonance lock at the observed $\dot{P}$ (even during the rapid evolutionary stage just before the convective core disappears).    In the future, we plan to investigate whether resonance locking is important in other short-period exoplanetary systems.

\subsection{Tidal dissipation on the main sequence}\label{sec:main_sequence}

We find that tidal dissipation on the main sequence is too inefficient to explain the observed $\dot{P}$.  Using the GYRE pulsation code \citep{Townsend:13} to solve the non-adiabatic oscillation equations for the WASP-12 models, we find that $\gamma t_{\rm gr} \approx 10^{-6}$ for internal gravity waves resonant with the tidal forcing.  Radiative damping is therefore an insignificant source of dissipation.  This is consistent with the results of \citet{Chernov:17}, who also consider radiative damping of the dynamical tide in main sequence models of WASP-12.  Although they show that the observed $\dot{P}$ could be explained if $\gamma t_{\rm gr}\sim 1$, which they refer to as the moderately large damping regime, they do not identify any mechanism that could enable the tide to be in this regime.  

Furthermore, we find $k_r \xi_r \ll 1$ throughout the propagation cavity of the main sequence models.  By Eq. (\ref{eq:krxir}), $k_r \xi_r$ is largest near the inner turning point, which for the main-sequence models  is located at $r\simeq 0.1 R_\odot$ (the top of the convective core); at this radius $k_r\xi_r \ll 1$. Thus, the dynamical tide does not break while the star is on the main sequence. 

Even if $k_r\xi_r \ll 1$ and the dynamical tide forms a standing wave, it can still potentially lose energy through weakly nonlinear interactions involving three-mode couplings  \citep{Essick:16}.  To check this, we computed three-mode coupling coefficients $\kappa_{abc}$ using the methods described in \citet{Weinberg:12}.  We considered the stability of the dynamical tide to the resonant parametric instability, which involves the tide (mode $a$) coupling to daughter $g$-modes (modes $b$ and $c$) whose eigenfrequencies satisfy $\omega_b + \omega_c \simeq \omega_a$.  We find that $\kappa_{abc}$ is small ($\kappa_{abc} \sim 1$ using the normalization in \citealt{Weinberg:12}) and the tide is stable to the parametric instability (i.e., the nonlinear growth rate $\Gamma < \gamma$).  We therefore conclude that while the star is on the main sequence, $\eta \ll 1$ and $P/|\dot{P}| \gg \trm{Myr}$.

\subsection{Tidal dissipation on the subgiant branch}\label{sec:subgiant}

In the subgiant models, the radiative damping rate $\gamma$ is again too small to significantly damp the dynamical tide. However, unlike the main sequence models, the subgiant models  have a {\it radiative} core and a convective envelope.  As a result, the inner turning point of the dynamical tide is much closer to the center of the star and we find that $k_r\xi_r \ga 1$ near the inner turning point.

Our numerical solutions give luminosities in the range  $L_{30}=[3.0, 10.5]$  for the subgiant models.  Specifically, for the subgiant model shown in Fig. \ref{fig:brunt_profile}, we find $L_{30}=3.6$.  For this model, the key parameters of the convective envelope are $r_c \simeq 1.30 R_\odot$, $M_c \simeq 1.20 M_\odot$, $\rho_c \simeq 2.3\times10^{-3} \trm{ g cm}^{-3}$ and the key parameters of the core are $\rho_2\simeq3.8$, and $C_{0.1}\simeq1.3$.  Plugging these into Eq. (\ref{eq:L}) and taking $M_p=1.40 M_{\rm Jup}$ \citep{Chan:11} gives  $L_{30}\simeq 16A_L$, which comparing to our numerical solution implies $A_L\simeq 0.2$.  Evaluating Eq. (\ref{eq:krxir}) at the inner turning point $r=\omega/C = 1.0\times 10^{-3} R_\odot$,  gives $k_r\xi_r =1.5$, in good agreement with the full numerical solution. Our other subgiant models yield very similar results, with values in the range $k_r\xi_r=[1.4,2.5]$.

This implies that during the subgiant phase, the dynamical tide becomes strongly nonlinear near the inner turning point and breaks.  As a result, $\eta \simeq 1$ and by Equation  (\ref{eq:PPdot}), the range in $L$ imply decay timescales in the range $P/|\dot{P}| = [1.4, 4.5]\trm{ Myr}$ (and $Q_\ast' =[0.8,2.2]\times10^5$).  This agrees well with the observed $P/|\dot{P}| = 3.2 \pm 0.3 \trm{ Myr}$.

Although we find $k_r \xi_r > 1$, it is only just slightly in excess of unity and one might wonder whether the wave really is efficiently damped ($\eta\simeq 1$).  Numerical simulations by \citet{Barker:11b} show  that as long as $k_r\xi_r > 1$, the wave breaks and efficiently transfers its angular momentum to the background mean flow.  Furthermore, \citet{Essick:16} find that if $k_r \xi_r \ga 0.1$, the dissipation due to weakly nonlinear interactions with secondary waves is nearly as efficient as when $k_r\xi_r \ga 1$.  This therefore suggests that $\eta \simeq 1$ for the WASP-12 subgiant models.

\section{Discussion}
\label{sec:discussion}

The main sequence and subgiant models are both $\approx 3\trm{ Gyr}$ old and spend $\approx 0.5\trm{ Gyr}$ within the measured range of $T_{\rm eff}$ and $\rho_\ast$.  If the observed $\dot{P}$ is indeed due to orbital decay, then an advantage of the subgiant scenario is that it naturally explains why the planet survived for 3~Gyr and is now decaying on a 3~Myr timescale.  Although the system only spends $\sim 0.1 \%$ of its life in the present state,  there are $\simeq 30$ hot Jupiters with $P< 3\trm{ days}$ orbiting  stars with $M_\ast> 1.2 M_\odot$.  The dynamical tide likely breaks during the subgiant phase in these systems and thus they spend $\sim 0.1\%$ of their $\sim$Gyr long lives in a state during which the planet decays on $\sim$Myr timescales.\footnote{As an aside, we note that because $P/|\dot{P}|$ increases rapidly with $P$, this mechanism cannot explain the apparent deficit of giant planets orbiting subgiants with periods between 10 and 100 days discussed in \citet{Schlaufman:13}.} We therefore estimate that out of the 30 systems, the probability of detecting one in a state like WASP-12 is $\sim 3\%$.

Even though wave breaking of the dynamical tide can drive orbital decay on Myr timescales, it cannot spin up and align the entire star.  This is because the wave breaks very close to the stellar center ($r< 0.01 R_\odot$) and while the torque $L/\Omega$ might spin up the stellar core \citep{Barker:10}, it is  too small to strongly affect the spin at the stellar surface. Therefore, our results do not conflict with the observed slow, misaligned rotation of WASP-12.

A combination of continued transit timing and occultation observations over the next few years should resolve whether the WASP-12 timing anomalies are due to orbital decay or apsidal precession \citep{Patra:17}. Since we find that the decay scenario is only plausible if the star is a subgiant, tighter constraints on the stellar parameters can also help provide resolution.  Given stellar modeling uncertainties, better constraints on just $T_{\rm eff}$ and $\rho_\ast$ might not be sufficient.  Asterosesimology offers a promising alternative.  Asteroseismic studies have determined whether stars are subgiants by measuring the sizes of convective cores \citep{Deheuvels:16} and measured the mass and radii of stars hosting planets to few percent accuracy \citep{Huber:13}.

\vspace{-0.3cm}
\acknowledgements{We thank Kishore Patra and Josh Winn for useful conversations and the referee for comments that improved the paper.  This work was supported by NASA grant NNX14AB40G.}
\vspace{-0.2cm}

\begin{thebibliography}{}
\makeatletter
\relax
\def\mn@urlcharsother{\let\do\@makeother \do\$\do\&\do\#\do\^\do\_\do\%\do\~}
\def\mn@doi{\begingroup\mn@urlcharsother \@ifnextchar [ {\mn@doi@}
  {\mn@doi@[]}}
\def\mn@doi@[#1]#2{\def\@tempa{#1}\ifx\@tempa\@empty \href
  {http://dx.doi.org/#2} {doi:#2}\else \href {http://dx.doi.org/#2} {#1}\fi
  \endgroup}
\def\mn@eprint#1#2{\mn@eprint@#1:#2::\@nil}
\def\mn@eprint@arXiv#1{\href {http://arxiv.org/abs/#1} {{\tt arXiv:#1}}}
\def\mn@eprint@dblp#1{\href {http://dblp.uni-trier.de/rec/bibtex/#1.xml}
  {dblp:#1}}
\def\mn@eprint@#1:#2:#3:#4\@nil{\def\@tempa {#1}\def\@tempb {#2}\def\@tempc
  {#3}\ifx \@tempc \@empty \let \@tempc \@tempb \let \@tempb \@tempa \fi \ifx
  \@tempb \@empty \def\@tempb {arXiv}\fi \@ifundefined
  {mn@eprint@\@tempb}{\@tempb:\@tempc}{\expandafter \expandafter \csname
  mn@eprint@\@tempb\endcsname \expandafter{\@tempc}}}

\bibitem[\protect\citeauthoryear{{Albrecht} et~al.,}{{Albrecht}
  et~al.}{2012}]{Albrecht:12}
{Albrecht} S.,  et~al., 2012, \mn@doi [\apj] {10.1088/0004-637X/757/1/18},
  \href {http://adsabs.harvard.edu/abs/2012ApJ...757...18A} {757, 18}

\bibitem[\protect\citeauthoryear{{Applegate}}{{Applegate}}{1992}]{Applegate:92}
{Applegate} J.~H.,  1992, \mn@doi [\apj] {10.1086/170967}, \href
  {http://adsabs.harvard.edu/abs/1992ApJ...385..621A} {385, 621}

\bibitem[\protect\citeauthoryear{{Arras} \& {Socrates}}{{Arras} \&
  {Socrates}}{2010}]{Arras:10}
{Arras} P.,  {Socrates} A.,  2010, \mn@doi [\apj] {10.1088/0004-637X/714/1/1},
  \href {http://adsabs.harvard.edu/abs/2010ApJ...714....1A} {714, 1}

\bibitem[\protect\citeauthoryear{{Asplund}, {Grevesse}, {Sauval}  \&
  {Scott}}{{Asplund} et~al.}{2009}]{Asplund:09}
{Asplund} M.,  {Grevesse} N.,  {Sauval} A.~J.,   {Scott} P.,  2009, \mn@doi
  [\araa] {10.1146/annurev.astro.46.060407.145222}, \href
  {http://adsabs.harvard.edu/abs/2009ARA%26A..47..481A} {47, 481}

\bibitem[\protect\citeauthoryear{Barker}{Barker}{2011}]{Barker:11b}
Barker A.~J.,  2011, \mn@doi [\mnras] {10.1111/j.1365-2966.2011.18468.x}, 414,
  1365

\bibitem[\protect\citeauthoryear{{Barker} \& {Ogilvie}}{{Barker} \&
  {Ogilvie}}{2010}]{Barker:10}
{Barker} A.~J.,  {Ogilvie} G.~I.,  2010, \mn@doi [\mnras]
  {10.1111/j.1365-2966.2010.16400.x}, \href
  {http://adsabs.harvard.edu/abs/2010MNRAS.404.1849B} {404, 1849}

\bibitem[\protect\citeauthoryear{Birkby et~al.,}{Birkby
  et~al.}{2014}]{Birkby:14}
Birkby J.~L.,  et~al., 2014, \mn@doi [\mnras] {10.1093/mnras/stu343}, 440, 1470

\bibitem[\protect\citeauthoryear{{Burkart}, {Quataert}, {Arras}  \&
  {Weinberg}}{{Burkart} et~al.}{2013}]{Burkart:13}
{Burkart} J.,  {Quataert} E.,  {Arras} P.,   {Weinberg} N.~N.,  2013, \mn@doi
  [\mnras] {10.1093/mnras/stt726}, \href
  {http://adsabs.harvard.edu/abs/2013MNRAS.433..332B} {433, 332}

\bibitem[\protect\citeauthoryear{{Burkart}, {Quataert}  \& {Arras}}{{Burkart}
  et~al.}{2014}]{Burkart:14}
{Burkart} J.,  {Quataert} E.,   {Arras} P.,  2014, \mn@doi [\mnras]
  {10.1093/mnras/stu1366}, \href
  {http://adsabs.harvard.edu/abs/2014MNRAS.443.2957B} {443, 2957}

\bibitem[\protect\citeauthoryear{Chan, Ingemyr, Winn, Holman, Sanchis-Ojeda,
  Esquerdo  \& Everett}{Chan et~al.}{2011}]{Chan:11}
Chan T.,  Ingemyr M.,  Winn J.~N.,  Holman M.~J.,  Sanchis-Ojeda R.,  Esquerdo
  G.,   Everett M.,  2011, \mn@doi [\aj] {10.1088/0004-6256/141/6/179}, 141,
  179

\bibitem[\protect\citeauthoryear{{Chernov}, {Ivanov}  \&
  {Papaloizou}}{{Chernov} et~al.}{2017}]{Chernov:17}
{Chernov} S.~V.,  {Ivanov} P.~B.,   {Papaloizou} J.~C.~B.,  2017, \mn@doi
  [\mnras] {10.1093/mnras/stx1234}, \href
  {http://adsabs.harvard.edu/abs/2017MNRAS.470.2054C} {470, 2054}

\bibitem[\protect\citeauthoryear{{Choi}, {Dotter}, {Conroy}, {Cantiello},
  {Paxton}  \& {Johnson}}{{Choi} et~al.}{2016}]{Choi:16}
{Choi} J.,  {Dotter} A.,  {Conroy} C.,  {Cantiello} M.,  {Paxton} B.,
  {Johnson} B.~D.,  2016, \mn@doi [\apj] {10.3847/0004-637X/823/2/102}, \href
  {http://adsabs.harvard.edu/abs/2016ApJ...823..102C} {823, 102}

\bibitem[\protect\citeauthoryear{{Deheuvels}, {Brand{\~a}o}, {Silva Aguirre},
  {Ballot}, {Michel}, {Cunha}, {Lebreton}  \& {Appourchaux}}{{Deheuvels}
  et~al.}{2016}]{Deheuvels:16}
{Deheuvels} S.,  {Brand{\~a}o} I.,  {Silva Aguirre} V.,  {Ballot} J.,  {Michel}
  E.,  {Cunha} M.~S.,  {Lebreton} Y.,   {Appourchaux} T.,  2016, \mn@doi [\aap]
  {10.1051/0004-6361/201527967}, \href
  {http://adsabs.harvard.edu/abs/2016A%26A...589A..93D} {589, A93}

\bibitem[\protect\citeauthoryear{{Enoch}, {Collier Cameron}, {Parley}  \&
  {Hebb}}{{Enoch} et~al.}{2010}]{Enoch:10}
{Enoch} B.,  {Collier Cameron} A.,  {Parley} N.~R.,   {Hebb} L.,  2010, \mn@doi
  [\aap] {10.1051/0004-6361/201014326}, \href
  {http://adsabs.harvard.edu/abs/2010A%26A...516A..33E} {516, A33}

\bibitem[\protect\citeauthoryear{{Essick} \& {Weinberg}}{{Essick} \&
  {Weinberg}}{2016}]{Essick:16}
{Essick} R.,  {Weinberg} N.~N.,  2016, \mn@doi [\apj]
  {10.3847/0004-637X/816/1/18}, \href
  {http://adsabs.harvard.edu/abs/2016ApJ...816...18E} {816, 18}

\bibitem[\protect\citeauthoryear{{Fossati} et~al.,}{{Fossati}
  et~al.}{2010}]{Fossati:10}
{Fossati} L.,  et~al., 2010, \mn@doi [\apj] {10.1088/0004-637X/720/1/872},
  \href {http://adsabs.harvard.edu/abs/2010ApJ...720..872F} {720, 872}

\bibitem[\protect\citeauthoryear{{Fuller} \& {Lai}}{{Fuller} \&
  {Lai}}{2012}]{Fuller:12}
{Fuller} J.,  {Lai} D.,  2012, \mn@doi [\mnras]
  {10.1111/j.1365-2966.2011.20237.x}, \href
  {http://adsabs.harvard.edu/abs/2012MNRAS.420.3126F} {420, 3126}

\bibitem[\protect\citeauthoryear{{Fuller}, {Luan}  \& {Quataert}}{{Fuller}
  et~al.}{2016}]{Fuller:16}
{Fuller} J.,  {Luan} J.,   {Quataert} E.,  2016, \mn@doi [\mnras]
  {10.1093/mnras/stw609}, \href
  {http://adsabs.harvard.edu/abs/2016MNRAS.458.3867F} {458, 3867}

\bibitem[\protect\citeauthoryear{{Fuller}, {Hambleton}, {Shporer}, {Isaacson}
  \& {Thompson}}{{Fuller} et~al.}{2017}]{Fuller:17a}
{Fuller} J.,  {Hambleton} K.,  {Shporer} A.,  {Isaacson} H.,   {Thompson} S.,
  2017, preprint, \href {http://adsabs.harvard.edu/abs/2017arXiv170605053F} {}
  (\mn@eprint {arXiv} {1706.05053})

\bibitem[\protect\citeauthoryear{{Goldreich} \& {Nicholson}}{{Goldreich} \&
  {Nicholson}}{1989}]{Goldreich:89}
{Goldreich} P.,  {Nicholson} P.~D.,  1989, \mn@doi [\apj] {10.1086/167665},
  \href {http://adsabs.harvard.edu/abs/1989ApJ...342.1079G} {342, 1079}

\bibitem[\protect\citeauthoryear{{Goldreich} \& {Soter}}{{Goldreich} \&
  {Soter}}{1966}]{Goldreich:66}
{Goldreich} P.,  {Soter} S.,  1966, \mn@doi [Icarus]
  {10.1016/0019-1035(66)90051-0}, \href
  {http://adsabs.harvard.edu/abs/1966Icar....5..375G} {5, 375}

\bibitem[\protect\citeauthoryear{{Goodman} \& {Dickson}}{{Goodman} \&
  {Dickson}}{1998}]{Goodman:98}
{Goodman} J.,  {Dickson} E.~S.,  1998, \mn@doi [\apj] {10.1086/306348}, \href
  {http://adsabs.harvard.edu/abs/1998ApJ...507..938G} {507, 938}

\bibitem[\protect\citeauthoryear{{Hansen}}{{Hansen}}{2010}]{Hansen:10}
{Hansen} B.~M.~S.,  2010, \mn@doi [\apj] {10.1088/0004-637X/723/1/285}, \href
  {http://adsabs.harvard.edu/abs/2010ApJ...723..285H} {723, 285}

\bibitem[\protect\citeauthoryear{Hebb et~al.,}{Hebb et~al.}{2009}]{Hebb:09}
Hebb L.,  et~al., 2009, \mn@doi [\apj] {10.1088/0004-637X/693/2/1920}, 693,
  1920

\bibitem[\protect\citeauthoryear{{Huber} et~al.,}{{Huber}
  et~al.}{2013}]{Huber:13}
{Huber} D.,  et~al., 2013, \mn@doi [\apj] {10.1088/0004-637X/767/2/127}, \href
  {http://adsabs.harvard.edu/abs/2013ApJ...767..127H} {767, 127}

\bibitem[\protect\citeauthoryear{{Husnoo}, {Pont}, {Mazeh}, {Fabrycky},
  {H{\'e}brard}, {Bouchy}  \& {Shporer}}{{Husnoo} et~al.}{2012}]{Husnoo:12}
{Husnoo} N.,  {Pont} F.,  {Mazeh} T.,  {Fabrycky} D.,  {H{\'e}brard} G.,
  {Bouchy} F.,   {Shporer} A.,  2012, \mn@doi [\mnras]
  {10.1111/j.1365-2966.2012.20839.x}, \href
  {http://adsabs.harvard.edu/abs/2012MNRAS.422.3151H} {422, 3151}

\bibitem[\protect\citeauthoryear{{Jackson}, {Greenberg}  \& {Barnes}}{{Jackson}
  et~al.}{2008}]{Jackson:08}
{Jackson} B.,  {Greenberg} R.,   {Barnes} R.,  2008, \mn@doi [\apj]
  {10.1086/529187}, \href {http://adsabs.harvard.edu/abs/2008ApJ...678.1396J}
  {678, 1396}

\bibitem[\protect\citeauthoryear{Jackson, Barnes  \& Greenberg}{Jackson
  et~al.}{2009}]{Jackson:09}
Jackson B.,  Barnes R.,   Greenberg R.,  2009, \mn@doi [\apj]
  {10.1088/0004-637X/698/2/1357}, \href
  {http://adsabs.harvard.edu/abs/2009ApJ...698.1357J} {698, 1357}

\bibitem[\protect\citeauthoryear{{Kushnir}, {Zaldarriaga}, {Kollmeier}  \&
  {Waldman}}{{Kushnir} et~al.}{2017}]{Kushnir:17}
{Kushnir} D.,  {Zaldarriaga} M.,  {Kollmeier} J.~A.,   {Waldman} R.,  2017,
  \mn@doi [\mnras] {10.1093/mnras/stx255}, \href
  {http://adsabs.harvard.edu/abs/2017MNRAS.467.2146K} {467, 2146}

\bibitem[\protect\citeauthoryear{{Maciejewski}, {Errmann}, {Raetz}, {Seeliger},
  {Spaleniak}  \& {Neuh{\"a}user}}{{Maciejewski} et~al.}{2011}]{Maciejewski:11}
{Maciejewski} G.,  {Errmann} R.,  {Raetz} S.,  {Seeliger} M.,  {Spaleniak} I.,
   {Neuh{\"a}user} R.,  2011, \mn@doi [\aap] {10.1051/0004-6361/201016268},
  \href {http://adsabs.harvard.edu/abs/2011A%26A...528A..65M} {528, A65}

\bibitem[\protect\citeauthoryear{{Maciejewski} et~al.,}{{Maciejewski}
  et~al.}{2016}]{Maciejewski:16}
{Maciejewski} G.,  et~al., 2016, \mn@doi [\aap] {10.1051/0004-6361/201628312},
  \href {http://adsabs.harvard.edu/abs/2016A%26A...588L...6M} {588, L6}

\bibitem[\protect\citeauthoryear{{Magic}, {Weiss}  \& {Asplund}}{{Magic}
  et~al.}{2015}]{Magic:15}
{Magic} Z.,  {Weiss} A.,   {Asplund} M.,  2015, \mn@doi [\aap]
  {10.1051/0004-6361/201423760}, \href
  {http://adsabs.harvard.edu/abs/2015A%26A...573A..89M} {573, A89}

\bibitem[\protect\citeauthoryear{{Moore} \& {Garaud}}{{Moore} \&
  {Garaud}}{2016}]{Moore:16}
{Moore} K.,  {Garaud} P.,  2016, \mn@doi [\apj] {10.3847/0004-637X/817/1/54},
  \href {http://adsabs.harvard.edu/abs/2016ApJ...817...54M} {817, 54}

\bibitem[\protect\citeauthoryear{{Moravveji}, {Aerts}, {P{\'a}pics}, {Triana}
  \& {Vandoren}}{{Moravveji} et~al.}{2015}]{Moravveji:15}
{Moravveji} E.,  {Aerts} C.,  {P{\'a}pics} P.~I.,  {Triana} S.~A.,   {Vandoren}
  B.,  2015, \mn@doi [\aap] {10.1051/0004-6361/201425290}, \href
  {http://adsabs.harvard.edu/abs/2015A%26A...580A..27M} {580, A27}

\bibitem[\protect\citeauthoryear{{Moravveji}, {Townsend}, {Aerts}  \&
  {Mathis}}{{Moravveji} et~al.}{2016}]{Moravveji:16}
{Moravveji} E.,  {Townsend} R.~H.~D.,  {Aerts} C.,   {Mathis} S.,  2016,
  \mn@doi [\apj] {10.3847/0004-637X/823/2/130}, \href
  {http://adsabs.harvard.edu/abs/2016ApJ...823..130M} {823, 130}

\bibitem[\protect\citeauthoryear{Ogilvie}{Ogilvie}{2014}]{Ogilvie:14}
Ogilvie G.~I.,  2014, \mn@doi [Annual Review of Astronomy and Astrophysics]
  {10.1146/annurev-astro-081913-035941}, 52, 171

\bibitem[\protect\citeauthoryear{{Patra}, {Winn}, {Holman}, {Yu}, {Deming}  \&
  {Dai}}{{Patra} et~al.}{2017}]{Patra:17}
{Patra} K.~C.,  {Winn} J.~N.,  {Holman} M.~J.,  {Yu} L.,  {Deming} D.,   {Dai}
  F.,  2017, \mn@doi [\aj] {10.3847/1538-3881/aa6d75}, \href
  {http://adsabs.harvard.edu/abs/2017AJ....154....4P} {154, 4}

\bibitem[\protect\citeauthoryear{Paxton, Bildsten, Dotter, Herwig, Lesaffre  \&
  Timmes}{Paxton et~al.}{2011}]{Paxton:11}
Paxton B.,  Bildsten L.,  Dotter A.,  Herwig F.,  Lesaffre P.,   Timmes F.,
  2011, \mn@doi [\apjs] {10.1088/0067-0049/192/1/3}, 192, 3

\bibitem[\protect\citeauthoryear{Paxton et~al.,}{Paxton
  et~al.}{2013}]{Paxton:13}
Paxton B.,  et~al., 2013, \mn@doi [\apjs] {10.1088/0067-0049/208/1/4}, 208, 4

\bibitem[\protect\citeauthoryear{Paxton et~al.,}{Paxton
  et~al.}{2015}]{Paxton:15}
Paxton B.,  et~al., 2015, \mn@doi [\apjs] {10.1088/0067-0049/220/1/15}, 220, 15

\bibitem[\protect\citeauthoryear{{Penev} \& {Sasselov}}{{Penev} \&
  {Sasselov}}{2011}]{Penev:11}
{Penev} K.,  {Sasselov} D.,  2011, \mn@doi [\apj] {10.1088/0004-637X/731/1/67},
  \href {http://adsabs.harvard.edu/abs/2011ApJ...731...67P} {731, 67}

\bibitem[\protect\citeauthoryear{{Penev}, {Jackson}, {Spada}  \&
  {Thom}}{{Penev} et~al.}{2012}]{Penev:12}
{Penev} K.,  {Jackson} B.,  {Spada} F.,   {Thom} N.,  2012, \mn@doi [\apj]
  {10.1088/0004-637X/751/2/96}, \href
  {http://adsabs.harvard.edu/abs/2012ApJ...751...96P} {751, 96}

\bibitem[\protect\citeauthoryear{{Phinney}}{{Phinney}}{1992}]{Phinney:92}
{Phinney} E.~S.,  1992, \mn@doi [Philosophical Transactions of the Royal
  Society of London Series A] {10.1098/rsta.1992.0084}, \href
  {http://adsabs.harvard.edu/abs/1992RSPTA.341...39P} {341, 39}

\bibitem[\protect\citeauthoryear{{Rasio}, {Tout}, {Lubow}  \& {Livio}}{{Rasio}
  et~al.}{1996}]{Rasio:96}
{Rasio} F.~A.,  {Tout} C.~A.,  {Lubow} S.~H.,   {Livio} M.,  1996, \mn@doi
  [\apj] {10.1086/177941}, \href
  {http://adsabs.harvard.edu/abs/1996ApJ...470.1187R} {470, 1187}

\bibitem[\protect\citeauthoryear{{Schlaufman}}{{Schlaufman}}{2010}]{Schlaufman:10}
{Schlaufman} K.~C.,  2010, \mn@doi [\apj] {10.1088/0004-637X/719/1/602}, \href
  {http://adsabs.harvard.edu/abs/2010ApJ...719..602S} {719, 602}

\bibitem[\protect\citeauthoryear{Schlaufman \& Winn}{Schlaufman \&
  Winn}{2013}]{Schlaufman:13}
Schlaufman K.~C.,  Winn J.~N.,  2013, \mn@doi [\apj]
  {10.1088/0004-637X/772/2/143}, \href
  {http://adsabs.harvard.edu/abs/2013ApJ...772..143S} {772, 143}

\bibitem[\protect\citeauthoryear{{Seager} \& {Mall{\'e}n-Ornelas}}{{Seager} \&
  {Mall{\'e}n-Ornelas}}{2003}]{Seager:03}
{Seager} S.,  {Mall{\'e}n-Ornelas} G.,  2003, \mn@doi [\apj] {10.1086/346105},
  \href {http://adsabs.harvard.edu/abs/2003ApJ...585.1038S} {585, 1038}

\bibitem[\protect\citeauthoryear{{Silva Aguirre}, {Ballot}, {Serenelli}  \&
  {Weiss}}{{Silva Aguirre} et~al.}{2011}]{Silva:11}
{Silva Aguirre} V.,  {Ballot} J.,  {Serenelli} A.~M.,   {Weiss} A.,  2011,
  \mn@doi [\aap] {10.1051/0004-6361/201015847}, \href
  {http://adsabs.harvard.edu/abs/2011A%26A...529A..63S} {529, A63}

\bibitem[\protect\citeauthoryear{{Teitler} \& {K{\"o}nigl}}{{Teitler} \&
  {K{\"o}nigl}}{2014}]{Teitler:14}
{Teitler} S.,  {K{\"o}nigl} A.,  2014, \mn@doi [\apj]
  {10.1088/0004-637X/786/2/139}, \href
  {http://adsabs.harvard.edu/abs/2014ApJ...786..139T} {786, 139}

\bibitem[\protect\citeauthoryear{{Torres}, {Fischer}, {Sozzetti}, {Buchhave},
  {Winn}, {Holman}  \& {Carter}}{{Torres} et~al.}{2012}]{Torres:12}
{Torres} G.,  {Fischer} D.~A.,  {Sozzetti} A.,  {Buchhave} L.~A.,  {Winn}
  J.~N.,  {Holman} M.~J.,   {Carter} J.~A.,  2012, \mn@doi [\apj]
  {10.1088/0004-637X/757/2/161}, \href
  {http://adsabs.harvard.edu/abs/2012ApJ...757..161T} {757, 161}

\bibitem[\protect\citeauthoryear{{Townsend} \& {Teitler}}{{Townsend} \&
  {Teitler}}{2013}]{Townsend:13}
{Townsend} R.~H.~D.,  {Teitler} S.~A.,  2013, \mn@doi [\mnras]
  {10.1093/mnras/stt1533}, \href
  {http://adsabs.harvard.edu/abs/2013MNRAS.435.3406T} {435, 3406}

\bibitem[\protect\citeauthoryear{{Watson} \& {Marsh}}{{Watson} \&
  {Marsh}}{2010}]{Watson:10}
{Watson} C.~A.,  {Marsh} T.~R.,  2010, \mn@doi [\mnras]
  {10.1111/j.1365-2966.2010.16602.x}, \href
  {http://adsabs.harvard.edu/abs/2010MNRAS.405.2037W} {405, 2037}

\bibitem[\protect\citeauthoryear{{Weinberg}, {Arras}, {Quataert}  \&
  {Burkart}}{{Weinberg} et~al.}{2012}]{Weinberg:12}
{Weinberg} N.~N.,  {Arras} P.,  {Quataert} E.,   {Burkart} J.,  2012, \mn@doi
  [\apj] {10.1088/0004-637X/751/2/136}, \href
  {http://adsabs.harvard.edu/abs/2012ApJ...751..136W} {751, 136}

\bibitem[\protect\citeauthoryear{{Witte} \& {Savonije}}{{Witte} \&
  {Savonije}}{2002}]{Witte:02}
{Witte} M.~G.,  {Savonije} G.~J.,  2002, \mn@doi [\aap]
  {10.1051/0004-6361:20020155}, \href
  {http://adsabs.harvard.edu/abs/2002A%26A...386..222W} {386, 222}

\bibitem[\protect\citeauthoryear{{Zahn}}{{Zahn}}{1975}]{Zahn:75}
{Zahn} J.-P.,  1975, \aap, \href
  {http://adsabs.harvard.edu/abs/1975A%26A....41..329Z} {41, 329}

\bibitem[\protect\citeauthoryear{{Zahn}}{{Zahn}}{1977}]{Zahn:77}
{Zahn} J.-P.,  1977, \aap, \href
  {http://adsabs.harvard.edu/abs/1977A%26A....57..383Z} {57, 383}

\makeatother
\end{thebibliography}


\end{document}